\documentclass[journal=jacsat,manuscript=article]{achemso}

\usepackage[version=3]{mhchem} % Formula subscripts using \ce{}
\usepackage{xcolor}
\usepackage{soul}
\usepackage{xr}
\usepackage{cleveref}

\author{Rocco Duquennoy}
\affiliation[CNR-INO]{National Institute of Optics (CNR-INO), Via Nello Carrara 1, Sesto Fiorentino, 50019, Italy}
\altaffiliation{Physics Department - University of Naples, via Cinthia 21, Fuorigrotta 80126, Italy}
\altaffiliation{European Laboratory for Non-Linear Spectroscopy (LENS), Via Nello Carrara 1, Sesto F.no 50019, Italy}

\author{Simon Landrieux}
\affiliation[LCF]{Laboratoire Charles Fabry, Institut d’Optique Graduate School, CNRS, Université Paris-Saclay, 91127 Palaiseau, France}
\altaffiliation{European Laboratory for Non-Linear Spectroscopy (LENS), Via Nello Carrara 1, Sesto F.no 50019, Italy}

\author{Daniele De Bernardis}
\affiliation[CNR-INO]{National Institute of Optics (CNR-INO), Via Nello Carrara 1, Sesto Fiorentino, 50019, Italy}
\altaffiliation{European Laboratory for Non-Linear Spectroscopy (LENS), Via Nello Carrara 1, Sesto F.no 50019, Italy}

\author{Juergen Mony}
\affiliation[CNR-INO]{National Institute of Optics (CNR-INO), Via Nello Carrara 1, Sesto Fiorentino, 50019, Italy}

\author{Maja Colautti}
\affiliation[CNR-INO]{National Institute of Optics (CNR-INO), Via Nello Carrara 1, Sesto Fiorentino, 50019, Italy}
\altaffiliation{European Laboratory for Non-Linear Spectroscopy (LENS), Via Nello Carrara 1, Sesto F.no 50019, Italy}

\author{Lin Jin}
\affiliation[KIP]{Kirchhoff-Institute for Physics, Heidelberg University, Im Neuenheim Feld 227, 69120 Heidelberg, Germany}
\altaffiliation{Institute of Physics, University of Münster, Wilhelm-Klemm-Strasse 10, 48149 Münster, Germany}

\author{Wolfram H.P. Pernice}
\affiliation[KIP]{Kirchhoff-Institute for Physics, Heidelberg University, Im Neuenheim Feld 227, 69120 Heidelberg, Germany}
\altaffiliation{Institute of Physics, University of Münster, Wilhelm-Klemm-Strasse 10, 48149 Münster, Germany}

\author{Costanza Toninelli}
\affiliation[CNR-INO]{National Institute of Optics (CNR-INO), Via Nello Carrara 1, Sesto Fiorentino, 50019, Italy}
\altaffiliation{European Laboratory for Non-Linear Spectroscopy (LENS), Via Nello Carrara 1, Sesto F.no 50019, Italy}

\email{toninelli@lens.unifi.it}

\title[An \textsf{achemso} demo]
  {Enhanced control of single-molecule emission frequency and spectral diffusion}
\abbreviations{IR,NMR,UV}

\begin{document}

%%%%%%%%%%%%%%%%%%%%%%%%%%%%%%%%%%%%%%%%%%%%%%%%%%%%%%%%%%%%%%%%%%%%%
%% The "tocentry" environment can be used to create an entry for the
%% graphical table of contents. It is given here as some journals
%% require that it is printed as part of the abstract page. It will
%% be automatically moved as appropriate.
%%%%%%%%%%%%%%%%%%%%%%%%%%%%%%%%%%%%%%%%%%%%%%%%%%%%%%%%%%%%%%%%%%%%%

%%%%%%%%%%%%%%%%%%%%%%%%%%%%%%%%%%%%%%%%%%%%%%%%%%%%%%%%%%%%%%%%%%%%%
%% The abstract environment will automatically gobble the contents
%% if an abstract is not used by the target journal.
%%%%%%%%%%%%%%%%%%%%%%%%%%%%%%%%%%%%%%%%%%%%%%%%%%%%%%%%%%%%%%%%%%%%%
\begin{abstract}
The Stark effect provides a powerful  method to shift the spectra of molecules, atoms and electronic transitions in general, becoming one of the simplest and most straightforward way to tune the frequency of quantum emitters by means of a static electric field. At the same time, in order to reduce the emitter sensitivity to charge noise, inversion symmetric systems are typically designed, providing a stable emission frequency, with a quadratic-only dependence on the applied field. However, such nonlinear behaviour might reflect in correlations between the tuning ability and unwanted spectral fluctuations. Here, we provide experimental evidence of this trend, using molecular quantum emitters in the solid state cooled down to liquid helium temperatures. We finally combine the electric field generated by electrodes, which results parallel to the molecule induced dipole, to optically excite long-lived charge states, acting in the perpendicular direction. Based on the anisotropy of the molecule's polarizability, our two-dimensional control of the local electric field allows not only to tune the emitter's frequency but also to sensibly suppress the spectral instabilities associated to field fluctuations.
% The quadratic dependence of the frequency shift from the applied field increases the sensitivity of the emitter to the electric field noise, constituting a dangerous backfire that degrades the photostability of the system.
%Here we experimentally investigate the Stark shift as a control tool on a organic molecule quantum emitter setup.
%Combining the electric field generated by parallel electrodes to the electric field due to optically-transported charges in the perpendicular direction, we demonstrate high degree of tunability of the molecular zero-phonon line.
%Thanks to the inequivalence of the molecule's polarizability and field fluctuations along different axis, this two-dimensional control of the local electric field allows not only to tune the emitter's frequency but also to sensibly suppress its spectral diffusion.

\end{abstract}
\section{Introduction}

Photons have a central role in quantum technology as flying qubits in quantum communication implementations, \cite{zapatero2023, hu2023} as well as in quantum sensing and quantum metrology experiments \cite{polino2020, pirandola2018}.  Besides the recent advances in integrated platforms for the implementation of linear optical quantum computing and quantum simulations \cite{kok2007, maring2024}, quantum advantage has been demonstrated with photons in boson sampling experiments \cite{Zhong2020, madsen2022}. Most of these applications rely on two-particle interference, resulting in an effective interaction between photons, ideally requiring indistinguishable photons on-demand \cite{Bouchard2021}.

Several methods can be used to produce single photons from solid state platforms, such as spontaneous parametric down conversion (SPDC) \cite{Jin2021} or the radiative decay of triggered quantum emitters, like organic molecules\cite{Toninelli2021}, quantum dots\cite{Senellart2017}, or color centers in diamond \cite{Sipahigil2019}.
%Mass-produced with cost-effective methods from organic synthesis and tailored optical properties, polyaromathic hydrocarbon molecules in host matrices have shown great potential for photonic quantum technologies. In particular, single photons sources have been demonstrated with high-purity and Fourier-limited linewidths \cite{Pazzagli2018, Rezai2019, Rezai2018}. 
%These are also one of the few systems for which two-photon interference has been shown both upon demultiplexing from a single molecule \cite{Trebbia2010, Rezai2018, Lombardi2021} and by combining the emission from independent sources \cite{Lettow2010a, Duquennoy2022}. 
Exploiting quantum emitters is desirable to overcome the hurdles of probabilsitic generation processes. Nevertheless, residual spectral diffusion critically prevents scaling up the number of photons using independent sources. Besides a tuning method able to compensate for the inhomogeneous broadening of the ensemble, this case requires indeed a high level of absolute frequency stability, as the emitters should remain resonant over the time scale of the entire experiment (typically for minutes).

The Stark effect provides a basic tuning mechanism which is available in most systems. This consists in imposing a controlled static electric field, that shifts by different amounts the energy levels involved in the transition \cite{Wild1992}.
However, it was demonstrated for instance with color centers in diamond and quantum dots, that the Stark effect increases the sensitivity of the emitter to the external electric field variations \cite{DeSantis2021, Conradt2023, Zuber2023, Gorlitz2022}. This results in random fluctuations of the emitter's line over multiple time scales, corresponding to the phenomenon known as spectral diffusion (SD)\cite{Kubo1969}. 

High single-photon purity, photostability and Fourier-limited linewidths are regularly measured in the emission from polyaromathic hydrocarbon molecules in host matrices \cite{Pazzagli2018, Rezai2019, Rezai2018}, making them well-established systems for photonic quantum technologies. Using an optically-induced Stark shift (OSS)  on dibenzoterrylene molecules in anthracene nanocrystals (DBT:Ac NCs) \cite{Colautti2020b}, quantum interference was observed in the emission from distinct molecules. In this experiment, the limited  visibility was ascribed to residual frequency fluctuations \cite{Lettow2010a, Duquennoy2022, Duquennoy2023}.It is thus a prominent challenge in photonic quantum technologies the tuning of quantum emitters, and among them single molecules, without backfire on their frequency stability.

%an all optical tuning mechanism observed in DBT:Ac emitters alongside other organic molecules based sources \cite{colautti_laser-induced_2020}. 

%The tuning technique used in these experiments relies on the Stark effect that describes the energy shift of the electronic energy levels due to the interaction of the system with the local electric field. \textbf{give references here}}  {\bf DDB note: aggiungere una parte su Stark shift in other devices like dots e color centers...}

In this paper we address this problem and demonstrate a viable path, by combining the aforementioned OSS technique with the more standard Stark tuning approach, due to electrodes-induced electric field. 
As a first result, we shed light on the unknown microscopic nature of the molecular-transition optical shift, by exploiting the externally applied electric field.
Then, a combination of optical pumping and electrodes-generated electric field is implemented (electrically-guided optical Stark shift:  EGOSS), yielding a high degree of control on the local applied field, with high spatial resolution and out of plane orientation. 
This allows for a reduction of spectral fluctuation (by a factor 12) for a given frequency shift, with respect to the case of an in-plane only electric field manipulation. 
In this way, the emission frequency of single molecules can be tuned without degrading their spectral stability. This paper hence suggests new understanding and crucial control tools for this widely common effect in solid state quantum emitters.

The article is organized as follows. In the first section we introduce the motivations and the main ideas behind the performed experiments. In the second one, 'Results and discussions', we expose the experimental results divided in three subsections. First, the Stark shift is presented for the zero-phonon line (ZPL) of DBT molecules due to the application of an external in-plane static electric field along the direction of the DBT principal axis, highlighting its impact on the molecule spectral diffusion. In the following discussion, the Stark effect from the externally applied electric field is combined with the OSS and the phenomenology of the EGOSS approach is characterized.
Finally, we demonstrate that EGOSS can be used to control the local electric field on the molecule allowing to tune its ZPL, while minimizing spectral diffusion.
All the setup and experimental details are reported in the 'Experimental Methods' section.
Further details regarding the theory and fitting procedure are reported in the 'Supporting Information' (SI).

\section{Motivations and basic ideas}
\label{sec:theory_intro}

Organic molecules are generally characterized by a single-photon electronic transition with a large dipole moment \cite{Toninelli2021}, which is dubbed 0-0 zero-phonon line (ZPL), as it does not couple to molecular vibrations. In particular, this transition is selected for the generation of indistinguishable photons, being lifetime-limited at cryogenic temperature.
As any purely electronic transition, the frequency of the ZPL can be modified by the presence of a static electric field based on the the well known Stark effect acting on the involved energy levels. The induced Stark shift can be expressed as follows:
\begin{equation}
    \hbar \Delta \omega_{\rm ZPL} =  - \vec{d}\cdot \vec{E}  - \vec{E}\cdot \boldsymbol{\alpha}\cdot \vec{E}.
\end{equation}
Here $\vec{E}=(E_x, E_y, E_z)$ is the static external electric field, $\vec{d}$  and $\boldsymbol{\alpha}$ are the difference in the static dipole moment and in the polarizability tensor between the ground and the excited state of the ZPL transition, respectively. For centrosymmetric molecules, $\vec{d}$ is expected to be zero and the Stark effect is completely dominated by the polarizability tensor difference $\boldsymbol{\alpha}$, giving a quadratic shift in the electric field strength. The polarizability tensor can be calculated by means of perturbation theory as discussed in the SI.
%The energy $\hbar \omega_{i}$ of the $i^{th}$ electronic level of the molecule is shifted by the presence of an electric field by a quantity $\hbar \Delta \omega_i = -\Vec{d}_i\cdot\Vec{E} + \Vec{E}\cdot\boldsymbol{\alpha}_i\cdot\Vec{E}$ where $\Vec{E}$ is the applied electric field and $\Delta \omega_i$, $\Vec{d}_i$ and $\boldsymbol{\alpha}_i$ are respectively the frequency shift, the static electric dipole and the static polarizability tensor of the $i^{th}$ level. 

\begin{figure}
    \centering
    \includegraphics[width=0.7\textwidth]{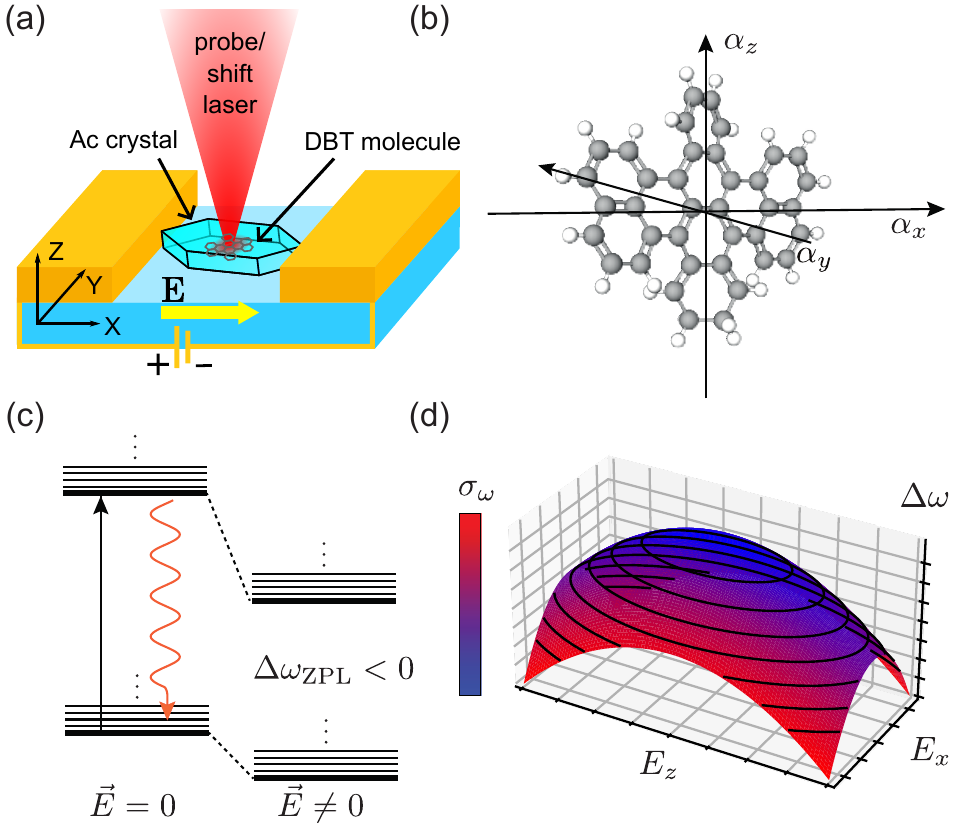}
    \caption{ (a) Sketch of the studied sample. DBT doped nanocrystals are dispersed on a glass substrate, nanostructured with $4\mu$m-gap interdigitated gold electrodes. The applied voltage can be scanned in a [-100:+100] V range, producing an electric field along the x direction. Optical access is available along the z axis and is used in excitation, for optical shift and fluorescence collection. (b) A cartoon picture of the DBT molecule, indicating the three symmetry axis corresponding to the diagonal components of the polarizability tensor. (c) Level structure of the ZPL before and after the application of a local electric field. The measurements are performed by resonantly exciting the ZPL and then detecting the Stokes-shifted emitted light, due to relaxation through a phononic side-band. (d) Illustrative example of the ZPL frequency shift as a function of the local electric field. The resulting surface is a squeezed paraboloid, narrower along the axis defined by $\alpha_{xx}$ and broader along the axis defined by $\alpha_{zz}$. The color scale represents the spectral diffusion experienced by the emitter increasing linearly with the value of the local electric field $(E_x, E_z)$. The contour lines represent the isofrequency curves. From this representation it is clear that it is possible to change SD by moving on an isofrequency line.
    }
    \label{fig:1}
\end{figure}

In this work, the molecular quantum emitter consists of DBT molecules inserted as impurities in Ac nanocrystals. The crystals are dispersed on glass between interdigitated electrodes, that are used to apply a controlled electric field (see Fig. \ref{fig:1}(a) for a sketch). The ZPL of the quantum emitter is given by the transition between the singlet excited state $|e\rangle$ and the ground state $|g\rangle$ with a characteristic frequency of $\omega_{\rm ZPL}/(2\pi) \approx 381$ THz. 
It is characterized by a high quantum yield $QY > 50\%$ \cite{Ren2022} and Debye-Waller factor $DW \simeq 70\%$ \cite{Clear2020}. At cryogenic temperatures ($T\approx 2.7 K$), the linewidth narrows down close to the Fourier-limit of $40$ MHz\cite{Duquennoy2022}.
Considering the geometry of the DBT molecule, mostly preserved after the insertion in the host matrix, we identify the three symmetry axis that diagonalize $\boldsymbol{\alpha}$, labelled as $x,y,z$ and reported in Fig.\ref{fig:1}(b). 
In this reference frame, the shift can be written as
\begin{equation}\label{eq:StarkShift_ZPL}
    \hbar \Delta \omega_{\rm ZPL} \approx -\alpha_{xx} E_x^2 - \alpha_{zz} E_z^2.
\end{equation}
Here we have neglected the $y$-component of the polarizability, $\alpha_{yy}$, which is expected to be much smaller than the other two, given the planar geometry of the molecule. 
%Indeed, ss much as the dipole moment, also the polarizability is limited by spatial extent, giving an intuitive explanation why in this case is neglected it, given the very thin size of the DBT in the $y$-direction.

As a distinct feature of the DBT molecule, its polarizability tensor difference is always positive along the $x$ and $z$ directions, leading to a red-shift quadratic in the field \cite{Nicolet2007b}, as schematically represented in Fig. \ref{fig:1}(c).
As a consequence, the Stark shift can be represented as a two-dimensional concave paraboloid as a function of the applied electric field, as depicted in Fig. \ref{fig:1}(d).
\footnote{Individual molecules can in general deviate from centrosymmetric geometry because of external factors like localized charge impurities and geometrical defects in the neighboring crystal sites \cite{moradi2019, Smit2024}. Even though those phenomena could induce $\vec{d}$ different from zero, the discussion can be formally mapped to the following one, by inserting a non-zero value of the local field $\vec{E}_0$ in absence of the external one.} With this powerful technique, shifts as large as $400$ GHz \cite{Schaedler2019} have been reported, allowing for the observation of cooperative effects among pairs of resonant molecules\cite{Trebbia2022, Hettich2002a}.

Notably, the sensitivity of the ZPL emission frequency to the local electric field is expected to have consequences in terms of the stability of the source's emission, becoming detrimental for experiments and applications based on multiple resonant emitters. In particular, at large fields, the molecule will be more affected by fluctuations in the external electric field or electric field noise due to fluctuacting charges\cite{Kambs2018, Kubo1969, Shkarin2021}, leading to enhanced spectral diffusion, (color scale Fig.\ref{fig:1}(d)). Indeed, 
considering the electric field only along the $x$-direction, and assuming that this is Gaussianly distributed around the value $E_x$ with variance $\sigma_E$, the shift probability will also be a Gaussian $p(\delta \omega, \Delta \omega_{\rm ZPL}, \sigma_{\omega})$ centered in $\hbar \Delta\omega_{\rm ZPL} = -\alpha_{xx} E_x^2$ with variance
$\hbar \sigma_{\omega} = 2\alpha_{xx}E_x\sigma_E$,
(See SI for major details).
Interestingly, these relations can be used to link together the Stark shift to the SD as follows:
\begin{equation}\label{eq:sigma_vs_StarkShift}
    \hbar \sigma_{\omega} = 2\sqrt{\alpha_{xx} \sigma_E^2 \, \hbar \Delta \omega_{\rm ZPL}},
\end{equation}
explicitly showing the characteristic square-root dependence of the Gaussian line-broadening component with the frequency shift and the polarizability tensor component.
\footnote{In a more realistic description, independent contribution to spectral fluctuations that can arise from electric field noise in different directions or from other physical processes, would add up quadratically. }

In this work, we have taken further this description, experimentally testing the idea that one can apply electric fields along different directions and obtain the same frequency shift. Having the SD mainly dominated by charge noise, an eventual anisotropy in the polarizability tensor will result in a independent control of SD. Indeed, given the different values of the absolute Stark shift gradient along the molecule's isofrequency lines represented in Fig. \ref{fig:1}(d), one could apply the tuning electric field in the direction where the polarizability tensor is minimum, attaining the same frequency shift while minimizing the electric-field sensitivity.
The impact of the presented findings go beyond the field of molecular single photon sources, putting forward a possible strategy for performing quantum optics experiments with indistinguishable photons from multiple distinct emitters.

%The origin of SD is typically ascribed to the presence of charge noise at the emitter level that results in a fluctuating local electric field. 

\section{Results and discussions}
\label{sec:results&discussion}

\begin{figure}
    \centering
    \includegraphics[width=\textwidth]{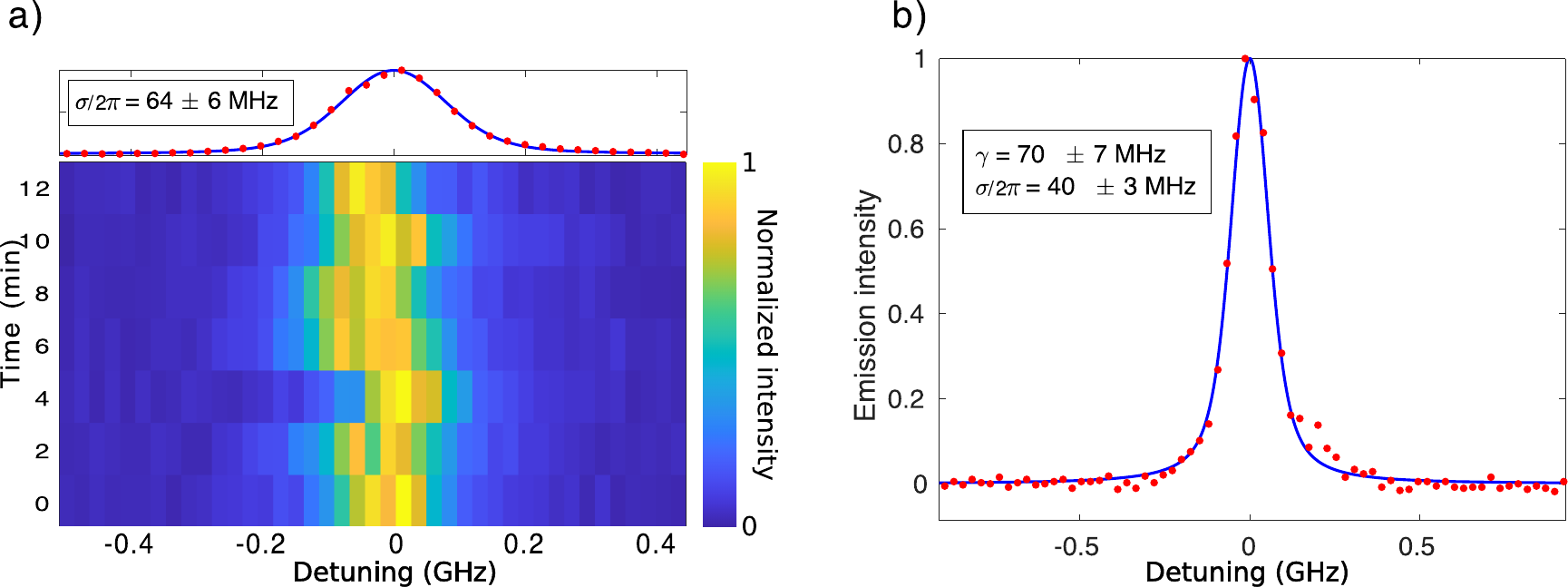}
    \caption{\textbf{Excitation spectroscopy and SD.} The excitation spectrum from a DBT emitter, obtained recording the red-shifted fluorescence signal (red dots) as a function of the excitation laser frequency is fitted by a Voigt profile (blue solid lines). The homogeneous ($\gamma$) and the inhomogeneous ($\sigma$) broadening components can be extracted from the fits. Here, each overall swipe is obtained at a scan speed of 0.5 GHz/s. (a) Each line corresponds to an excitation spectrum plotted in color scale. The total measurement time for the repeated swipes amounts to 12 minutes. The integrated signal is plotted in the top panel. (b) A single snap shot is shown where the lower $\sigma$ component confirms that the SD depends on the observation time.}
    \label{fig:2}
\end{figure}

In this section we describe the obtained experimental results, which well confirm the intuitive picture presented in the previous Section. The basic technique used in this work is excitation spectroscopy: scanning a tunable laser over the ZPL and collecting the Stokes-shifted fluorescence after spectral filtering, we measure the molecule excitation spectrum (color-coded traces in Fig.\ref{fig:2}(a)). For the purpose of this work, it will be sufficient to observe isolated lines to ascribe them to independent emitters.
Due to the fluctuating electric environment, these spectra diffuse in time, as it can be observed in the repeated set of measurements shown in Fig. \ref{fig:2}(a).
By integrating such measurements over time, a profile such as the one reported on top panel is obtained. 
This type of data yields  information about SD, by fitting a Voigt profile and extracting its gaussian component (for more details see the SI). 
In the case of Fig. \ref{fig:2}(a) for example, we estimate $\sigma_{\omega}/(2\pi) = 64 \pm 6$ MHz. Here, each line corresponds to a laser scan at 0.5 GHz/s scanning speed, and a $2$ min interval is to be considered among subsequent scans, resulting in an overall measurement time over $12$ minute. For a single trace, we obtain a characteristic profile like the one reported in Fig. \ref{fig:2}(b).
It is worth noticing that also in this case, the measured line is well fitted by a Voigt profile resulting from the convolution of a Lorentzian line with FWHM $\gamma/(2\pi) = 70 \pm 7$ MHz and a Gaussian distribution $p(\delta\omega)$ with $\sigma_{\omega}/(2\pi) = 40 \pm 3$ MHz. 
In fact, the observation time required to perform a single laser scan, typically around $3$ s, provides an intrinsic integration time for our measurements, during which SD cannot be completely neglected. Comparing the two profiles in Fig. \ref{fig:2}, it is clear that the width of the Gaussian component in the spectrum (which is associated to SD) critically depends on the observation time.

%The width of the Gaussian component of the spectrum (which we associate to SD) depends on the observation time as it can be shown by comparing the aforementioned results in Fig. \ref{fig:2} and the signal is integrated for an overall interrogation time of 12 minutes, (top panel).

%This methodology has been adopted to analyse SD throughout this whole experimental work.
Using this methodology we now investigate how the molecular emission properties are modified under the influence of an applied electric field.

\subsection{Principal-axis electric Stark shift and related spectral diffusion}
\label{sec:principal_axis_shift}

%The application of a static electric field generates a Stark shift of the energy levels of the emitter. In the electric dipole approximation the energy shift is:
%\textcolor{blue}{you have to add hbar, or write it in terms of energies which is maybe better not to confuse it with the emission frequency}
%\begin{equation}
%    \hbar \Delta \omega^i = -\Vec{d}^i\cdot\Vec{E} - \Vec{E}\cdot\boldsymbol{\alpha}^i\cdot\Vec{E}
%\end{equation}

%where $\Vec{E}$ is the applied electric field and $\Delta \omega^i$, $\Vec{d}^i$ and $\boldsymbol{\alpha}^i$ are respectively the frequency shift, the static electric dipole and the static polarizability tensor of the $i^{th}$ level. In the case of the DBT:Ac emitter, the centrosymmetric nature of the molecule (see Fig. 1b) implies that the source cannot have a static electric dipole \cite{does it make sense to cite a basic QM book?} and thus shows only the typical quadratic behavior. Scanning the voltage applied to the electrodes (from -100 to +100 V) and monitoring the ZPL frequency at each voltage we obtain results of the kind showed in Fig.1c. A parabolic fit allows for the extraction of the $(x,x)$ component of $\Vec{\alpha}^{ZPL} = \Vec{\alpha}^{e} - \Vec{\alpha}^{g}$ as the coefficient of the quadratic term in the fitted parabolas ($e$ and $g$ indicating the excited and the ground state of the ZPL transition).

\begin{figure}[h]
    \centering
    \includegraphics{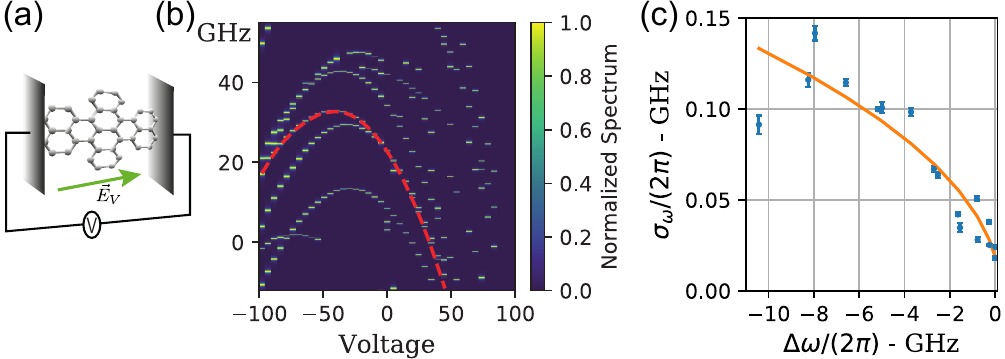}
    \caption{\textbf{Stark shift and spectral diffusion }(a) Cartoon picture of the specific configuration. We perform the spectral measurements on a DBT molecule embedded between two electrodes at fixed voltage difference generating a tunable in-plane electric field in the direction of the principal polarization axis defined by $\alpha_{xx}$.
    (b) Typical  measurement of the excitation spectra (color coded along the y axes of the figure) as a function of the applied voltage on the electrodes. Each parabola can be associated to a different molecule. The measurements are obtained with a scan speed of 0.5 GHz/s. At this scanning speed the observed line is broadened by the fast components of SD.
    (c) Measured SD as a function of the frequency shift extracted from the data highlighted by the red dashed line in (b). The values of $\sigma_{\omega}$ are extracted by fitting each fixed voltage spectrum with a Voigt profile. The orange line corresponds to the best fit of the square-root dependence of SD against shift predicted in Eq. \ref{eq:sigma_vs_StarkShift}.    }
    \label{fig:3}
\end{figure}

The Stark shift is measured here by applying a finite voltage  between the interdigitated electrodes shown in Fig.\ref{fig:1}(a), thus generating an in-plane electric field $\vec E_V$ along the $x$-direction, as reported in Fig. \ref{fig:3}(a). For each value of the electric field, an excitation spectrum is collected, and the fluorescence intensity is reported in color scale in Fig. \ref{fig:3}(b).
%The measurements of the Stark shift effect due to a static external electric field along the $x$-direction, generated by a pairs of electrodes are reported in Fig.\ref{fig:3}, (see experimental methods for further details).
As several DBT molecules are embedded in the same nanocrystal addressed in confocal microscopy, multiple lines appear in the figure columns, corresponding to the ZPL frequencies of the different emitters. As expected (see Eq. \eqref{eq:StarkShift_ZPL}), their frequency position shifts following a  parabolic behaviour as a function of the applied voltage. 
It is worth noticing that molecules do not  shift all by the same amount due to the applied electric field. This can be ascribed to a small disorder in the alignment with respect to b-axes \cite{Musavinezhad2024}. Moreover, the  local electric field difference giving rise to the inhomogeneous broadening, might also determine an effectively different response to the externally applied field.
For molecules whose dipole transition is perpendicular to the direction of the external electric field we do not detect any shift. This observation justifies the approximation $\alpha_{yy} = 0$ used in Eq. \eqref{eq:StarkShift_ZPL} and is coherent with the planar nature of the DBT molecule, for which charge mobility is strongly suppressed in the direction orthogonal of the molecular plane.
Indeed, as represented in Fig.\ref{fig:1}(a), the DBT molecule trapped in the Antracene crystal is expected to have the secondary axis in the $z$-direction, perpendicular to our substrate plane \cite{Nicolet2007b}. For these reasons, we expect that all the Stark effect observed due to the electrodes is given by the $\alpha_{xx}$ polarizability. The coefficient of the quadratic term of each parabola yields hence the $x$-polarizability difference of the molecule in the direction of the applied field. 
Since in our setup the electric field applied on the molecule through the electrodes can be as large as $E_x\approx \pm 200$kV/cm (see experimental methods), the polarizability along the $x$-direction can be estimated by fitting the parabolas in Fig. \ref{fig:3}(b) as $\alpha_{xx}/\hbar  \approx 2\pi \times 1.82 \pm 0.11 $MHz/(kV/cm)$^2$, in agreement with previous results \cite{Nicolet2007b}.
As discussed in the SI, this value is also reasonably close with the theoretical estimates for DBT \cite{Sadeq2018}, where  a range of values is reported $\alpha_{xx} / \hbar \approx 2\pi \times 0.2-2$MHz/(kV/cm)$^2$.

In the following we discuss the procedure to test the square root dependence of the spectral diffusion with the frequency shift. Each excitation spectrum is fitted with a Voigt profile, whose FWHM is given by \cite{Whiting1968} $2\pi \Gamma_{\rm Voigt} = \gamma_0/2 + \sqrt{(\gamma_0/2)^2 +  8\ln (2)\sigma_{\omega}^2  }$.
The Lorentzian component is fixed to $\gamma_{0}/(2\pi) = 80$ MHz, obtained as the linewidth of a Lorentzian fit to the narrowest measured line. From the Voigt profile, we hence extract the parameters $\sigma_{\omega}$, associated solely to SD.
The dependence of $\sigma_{\omega}$ on the measured Stark shift $\Delta \omega_{\rm ZPL}$ is found to be correctly expressed by Eq. \eqref{eq:sigma_vs_StarkShift} where an offset $\sigma_{0}$ is added quadratically, so as to take into account for charge noise in the $z$-direction. The results are shown in Fig.\ref{fig:3}(c), where  only measurements between -100 V and -15 V are considered, in order to avoid the region where the lines from two molecules were overlapping. From the fit to the data in Fig. \ref{fig:3}(c) with Eq. \ref{eq:sigma_vs_StarkShift}, the product of the electric field variance with the polarizability difference can be estimated, yielding a value $\alpha_{xx}\sigma_{E}^2 /\hbar = 2\pi \times 0.410 \pm 0.001$\, MHz. Using the previously estimated value for the polarizability difference, the local electric field variance can be obtained, amounting to a value $\sigma_E = 0.47\pm 0.20$\,kV/cm.
These electric field fluctuations can be probably ascribed to the presence of moving charges in the matrix environment where the molecule is hosted.
Regardless of the specific physical origin, it is interesting to compare this variance with the one of fluctuating fields in other material platforms. 
In this case, the quantum emitter is used as a probe to test the noise level of the electric environment where it is hosted.
For instance, this value is significantly smaller than the one recently measured in SnV centers\cite{DeSantis2021} and colloidal quantum dots \cite{Conradt2023}, whereas it is 3 orders of magnitude larger with respect to field fluctuations in charge-engineered semiconductor quantum dots \cite{Kuhlmann2013, Zhai2020, Zhai2022}. 
This observation puts the molecular platform in an excellent position for quantum optics experiment, as very high visibility of two-photon intereference have been observed also without charge engineering \cite{Lombardi2021}.

When the frequency shift is small, $\Delta \omega_{\rm ZPL} \rightarrow 0$, and the molecule is approaching the summit of its Stark-parabola, we observe that the SD decreases, approaching zero for certain molecules, such as the one shown in Fig. S.\ref{supp-fig:7} of the SI. The measurement time scale is here the sameone used for the data in Fig.\ref{fig:2}(b), corresponding to a scan speed of $0.5$ GHz/s.

\subsection{Two-dimensional electric field control: OSS and EGOSS}
\label{sec:secondary_axis_shift}

The laser-induced frequency shift of DBT emission has been discovered in recent years \cite{Colautti2020b} and is now an established tool in current quantum optics studies based on such molecular setups\cite{Duquennoy2022, Lange2024}.
The OSS technique is implemented focusing an intense light beam on the emitter at cryogenic temperatures, for periods of time of the order of minutes (1 to 16 minutes for the experiments here reported). The ZPLs of the embedded molecules shift to the red side of the spectrum and remains there even after the external laser has been switched off.
With respect to the more standard shifting approach using external electrodes, it has the advantage of being fabrication free and having a much higher spatial resolution, limited only by the diffraction limit of the shifting laser beam. Its reversibility is bounded by a typical relaxation time that is specific for every sample and typically take even a few hours\cite{Colautti2020b}.
These features come at the cost of being less reproducible.  
%While previous works on the optical shift technique offer a phenomenological characterization of the process and a possible modelling of it, experimental data on the microscopic nature of the phenomenon has not been reported yet.
\begin{figure}
    \centering
    \includegraphics[width=\textwidth]{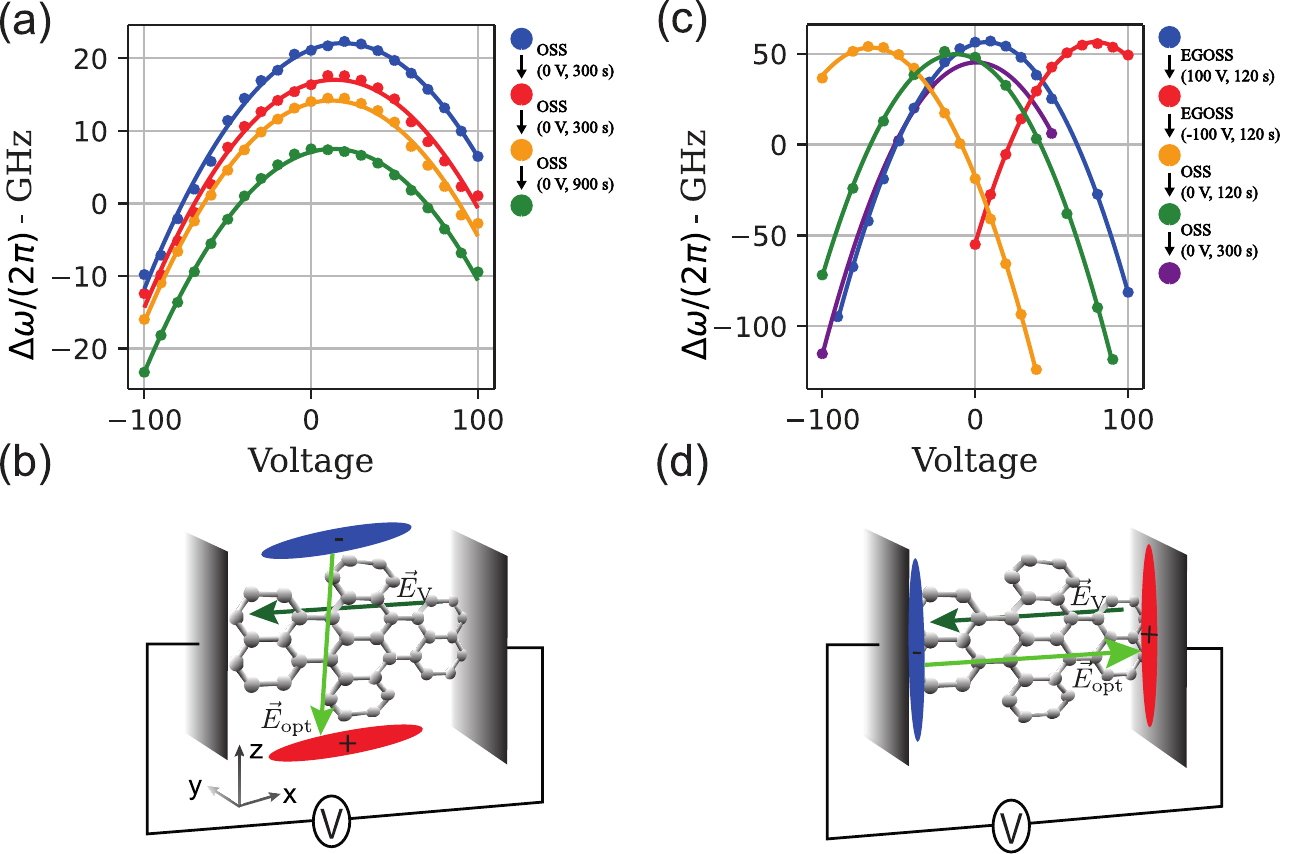}
    \caption{\textbf{OSS and EGOSS} (a) The measured ZPL frequency shift $\Delta \omega_{\rm ZPL}/(2\pi)$ after subsequent OSS runs are plotted as colored dots. Solid lines represent the best fit to the data with parabolic functions. The experimental configuration is sketched in panel (b). Applying the optical shift results in an accumulation of charges along the $z$-axis that gives rise to an electric field along the secondary polarizability axis of the DBT molecule; as a second step an in-plane electric field along the $x$-axis is applied through the electrodes, varying the voltage $V$.
    (c) The measured ZPL frequency shift $\Delta \omega_{\rm ZPL}/(2\pi)$ after subsequent EGOSS runs are plotted as colored dots. Solid lines represent the best fit to the data with parabolic functions. The experimental configuration is sketched in panel (d). In this case, fix bias voltages impose a preferential direction for the excited charges to move, accumulating them in the direction of the electrodes-applied field. This results in an effective screening of the in-plane electric field between the electrodes, hence in an electric field off-set in the Stark-parabola. All OSS and EGOSS cycles have been performed with 100$\mu$W pump power. Applied voltages and exposure times are reported in the legends next to the plots.}
    \label{fig:4}
\end{figure}
The laser-induced shift has been interpreted as a Stark shift induced by long-lived charge separated states arising from the ionization cascade induced by the laser illumination \cite{Colautti2020b}. These charges generate a local electric field, denoted here $\vec E_{\rm opt}$, which shifts the molecule frequency via Stark effect. 
However, the spatial distribution of these charges remains unknown, and previous studies do not provide any indication on the $\vec E_{\rm opt}$ direction. 

In the following we test this hypothesis and study the interplay of the optically-induced effect with the electrode-induced Stark shift, identifying a combined tuning method, i.e. the electrically guided OSS, hereafter named EGOSS.

%As shown in the following, our observations confirm the electric nature of the OSS, and demonstrate that a significant portion of the excited charges accumulate along the $z$-direction, likely following the symmetry of the Anthracene crystal structure.  
%Moreover, we explicitly show that the displaced charges can be redistributed using external electrodes.

At first, several runs of OSS are applied, each corresponding to a laser pulse with 10 kW/cm$^2$ intensity and exposure times reported in the legend of Fig. \ref{fig:4}(a).  After each one, the excitation spectrum is measured for different applied voltages along the x-axis and a characteristic parabola is reconstructed, by plotting the corresponding central frequency of the Voigt profile.  The effect of the OSS on the charge distribution can be appreciated by observing the different colored parabolas in Fig. \ref{fig:4}(a)), as they progressively down-shift towards lower emission frequencies.
This rigid translation indicates that the laser-generated charges have produced a field in a direction orthogonal to the one probed by the electrodes geometry. In this picture, one can look at the red-shifting parabolas as different slices of the paraboloid in Fig.\ref{fig:1}(c), corresponding to various values of the field along the z-axes. Indeed, as discussed in the previous sections, due to the thin geometry of the DBT molecule along the $y$-direction, we expect it to have negligible polarizability on that axis $\alpha_{yy} \sim 0$. This implies that the measured Stark effect can be attributed to the $\vec E_{\rm opt}$ electric field having a finite amplitude along the $z$-direction.
%To understand it we need to first consider previous results on the shape of the nanocrystals: as reported from simulations in \cite{Nicolet2007b}, in contrast to many commonly adopted representations that picture the emitters on the x-z plane, the DBT molecule lays on the x-y plane, as depicted in our cartoon Fig. \ref{fig:1}(a).
Although we do not have direct access to the spatial distribution of charges displaced by the optical pump, this observation suggests that a significant portion of this charge accumulation occurs along the $z$-direction, as schematically illustrated in Fig. \ref{fig:4}(b). Once this process is complete, the charges are trapped and their configuration remains stationary over the typical timescales of the experiment. 

Now, the idea can be put forward to use an external electric field to better control such generated charges in the OSS scheme and combine the benefits of the two methods in an electrically-guided OSS (EGOSS).  Indeed, we immediately notice that the OSS is strongly influenced by the contextual application of an external electric field. When a bias voltage is maintained on the electrodes during the OSS run, a subsequent voltage sweep on the electrodes produces a result markedly different from what  previously described. In this case, the parabolas shift also in the horizontal direction, indicating that the OSS generates an offset electric field in the same direction as the one generated by the electrodes. The final result is a displacement of the vertex of the parabola towards the applied bias voltage, as shown in Fig. \ref{fig:4}(c). The application of the EGOSS sequence at +100 V between the blue and the red parabolas show that, with an exposure time of 120 s, there is an almost complete screening of the external field at the emitter's position. Subsequent EGOSS sequences confirm this behavior and show that the screening can also be saturated with longer exposure times, as it can be appreciated considering the last OSS sequence between the green and the purple parabola that lasted for 300 s.
%\ddb{This effect occurs because, in the absence of a bias field, charges spontaneously move to a new quasi-equilibrium state}, accumulating along the $z$-direction, \ddb{likely} following the symmetries of the Anthracene crystal structure. 
%Once this process is \ddb{complete}, the charges \ddb{become} trapped \ddb{and remain stationary} over the typical timescales of this experiment. 
This behavior can be explained by considering that when optical pumping is performed while an external electric field is present, the charges move in the direction of the field to compensate for it (see Fig. \ref{fig:4}(d)). We interpret the saturation behavior of the screening as a consequence of the fact that the lower the local field, the harder it is to free charges from the defects of the crystal, making the effect progressively less efficient.
We note in passing that this shielding effect associated to EGOSS is effective also when the applied voltage is zero. Indeed, the green parabola is shifted towards zero voltage with respect to the yellow one, as a result of a simple OSS procedure (see Fig.\ref{fig:4} caption). This effect is less visible in panel (a) as the local field for the pristine crystal is already quite close to zero. It is worth noticing this technique is reversible, as the parabolas can move in both directions, with only a minor residual vertical component of the shift. Notably, the reported observations  represent an experimental evidence of the OSS phenomenon electrical nature and provide and excellent knob to tackle the problem of Stark-shift induced spectral diffusion.

\subsection{EGOSS for the reduction of Spectral Diffusion}

As a major outcome of this work, in this section we demonstrate that the EGOSS-assisted tuning method  allows for much smaller SD with respect to the case of tuning by the traditional in-plane electric field, while keeping control of the operating voltage.
Indeed, as $\alpha_{zz} < \alpha_{xx}$, following Eq. \eqref{eq:sigma_vs_StarkShift}, we expect that for equal values of the frequency shift, the optical shift acting along the z axes should determine a lower SD. Moreover, thanks to the presence of the external electric field, the experiment can be performed at arbitrary values of the applied voltage. This instance can be crucial when multiple emitters have to be brought in resonance.
%This concept arises from the simplified one-dimensional model leading to Eq. \eqref{eq:sigma_vs_StarkShift}: increasing the field in one direction shifts the emission frequency but also heightens the molecule's sensitivity to field fluctuations, thereby increasing SD. In this study, we investigate whether shifts using fields in $x$ or $z$ directions have different effects on SD.

%Indeed by considering Eq. \eqref{eq:sigma_vs_StarkShift} for the two components of the electric field $E_x, E_z$, we can compute the ratio between the SD variances at equal shift $\Delta\omega_{\rm ZPL}(E_x) = \Delta\omega_{\rm ZPL}(E_z)$ obtaining
%\begin{equation}\label{eq:alpha ratio}
%    \frac{\sigma_{\omega}(E_z) }{\sigma_{\omega}(E_x)} = \frac{\sigma_{E_x}}{\sigma_{E_z}}  \sqrt{\frac{\alpha_{zz}}{\alpha_{xx}}} < 1 .
%\end{equation}
%\textcolor{green}{TO ADD: and assuming the same $\Delta\omega$ within the experimental errors and fluctuations of the electric field characterized by the same fluctuations $\sigma_{E}$,}

%Here we have assumed that the electric field fluctuations have comparable variances on both directions.
%Visually, we can exploit the combination of optical shift and electrodes Stark effect to freely move on the paraboloid shown in Fig.\ref{fig:1}(c), to target a shift of a certain magnitude $\Delta\omega_{\rm ZPL}$ along the direction that minimize the SD.

\begin{figure}[h]
    \centering
    \includegraphics[width=0.5\textwidth]{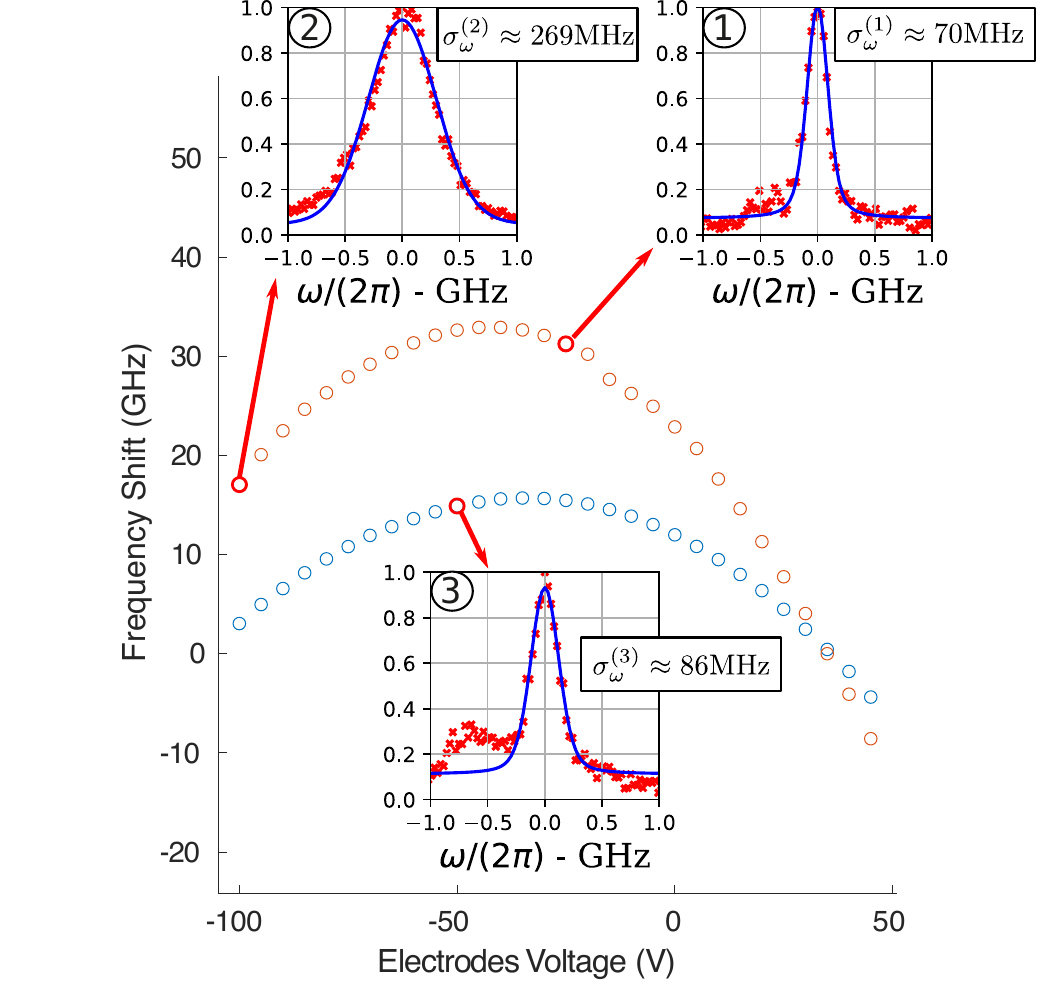}
    \caption{ \textbf{Effect of electric field control on SD.} The parabolas are obtained by monitoring the frequency of emission of a molecule during a scan over the electrodes voltage. The orange dots are measured before any optical shift while the blue ones after a combined shift that allowed for the creation of a local field in the $z$-direction. In correspondence to the red highlighted points we measured SD by repeating excitation spectroscopy measurements. The integrated signal is presented in the insets. A Voigt profile fit on each of the integrated signals allowed for the isolation of the gaussian component of the broadening associated to SD.
}
    \label{fig:5}
\end{figure}

We here consider for instance the case of a molecule that is intrinsically biased, hence exhibiting a parabola vertex at finite voltages (see e.g. orange dots in Fig.\ref{fig:5}). In order to compare the molecule SD at different emission frequencies, while operating the setup at similar values of the applied field, an EGOSS sequence at $-25$ V is applied. The following response to the in-plane electric field is shown as blue dots in  Fig. \ref{fig:5}. SD is then measured in correspondence to the three highlighted dots, averaging 30 excitation spectra (scan speed $0.5$ GHz/s) over a time span of about 12 minutes (red dots in  Fig.\ref{fig:5} insets).
%From the time traces, a small frequency drift common to multiple molecules can be observed. The drift can be ascribed to a long term drift of the frequency of emission of the exciting laser that can be approximately considered linear on the time scales of the measurements and thus is removed from the data before the integration of the signal. 
More details on the time traces are provided in the SI.
%The SD traces in the form of the one in Fig.\ref{fig:3}(c) are reported in Fig.\ref{fig:5}(b) while the integrated traces, each fitted with a Voigt profile, are shown in the insets of Fig.\ref{fig:5}(a).
Close to the vertex of the orange parabola, obtained before EGOSS, $\sigma_{\omega}^{(1)}$ amounts to a value of $ 70 \pm 2$ MHz. A 13-GHz shift obtained by applying an electrode-induced field results in an increased SD characterized by $\sigma_{\omega}^{(2)} = 269 \pm 3$ MHz. This significant stability degradation can be partially avoided exploiting the results presented in the previous section. The implementation of the EGOSS method leads to a frequency shift by a very similar amount (14 GHz), but with a much smaller increase of SD ($\sigma_{\omega}^{(3)} = 86 \pm 5$ MHz). Moreover, the tuning obtained with this combination is as persistent as the bare optical shift on the time scales of our experiments (hours). While the ratio of the post-shift SDs $\sigma_{\omega}^{(2)}/\sigma_{\omega}^{(3)} = 3.1$ already proves the effectiveness of the method, the relative increase with respect to the starting SD $(\sigma_{\omega}^{(2)}-\sigma_{\omega}^{(1)})/(\sigma_{\omega}^{(3)}-\sigma_{\omega}^{(1)}) = 12$ underlines how effective a higher degree of control on the fields can be in order to control SD.
The relation $\alpha_{xx} > \alpha_{zz}$ is expected from simple considerations on the size of the molecule in the $x$ and $z$ directions and is in line with the observed reduction of SD. However, such a big improvement suggests that also the field fluctuations $\sigma_{E}$ could be smaller in the vertical direction. Further measurements would be required in order to test this hypothesis.

\section{Conclusions}
\label{sec:conclusions}

In this work, we experimentally demonstrate 2-D control over the local electric field around a single DBT molecule, exploiting a combination of electrode-induced field and optically generated charge-separated states. These experiments enhance our understanding of the microscopic mechanisms behind the optically induced frequency shift (OSS) in molecules and allows for efficient control molecular emissions while remaining almost fabrication-free. 

By combining optical shifts with the electrode-induced field, we have developed a versatile method for shifting the ZPL, which is dubbed EGOSS. This dual approach allows to achieve significant shifts in the emission frequency, with minimal impact on the final frequency stability as opposed to ordinary approaches. The demonstrated improvement amounts to a factor 12, yielding a promising result for the advancement of organic molecules as single photon sources in quantum technologies. The methodologies used in this paper could also be exploited in future works to measure local charge noise in the environment of quantum emitters.

\section{Experimental methods}
\label{sec:exp_methods}

%\textcolor{red}{(Now this part is also in the introduction, shall I remove one of the two?) We here present our results on the topic obtained thanks to the implementation of an electrodes system that enables the application of a static electric field along the x direction on our sample (See Fig. 1) and that we use to probe the effect of the laser-induced shift on the local charge environment of a molecular quantum emitter. Next, we will show how, thanks to the acquired knowledge, a combination of both laser-induced and electrodes-induced techniques can be exploited in order to achieve a reduction in the spectral fluctuations of our emitters that could improve the level of indistinguishability for this kind of sources \cite{Duquennoy2022}}.

The sample used for these experiments consists of a glass substrate covered with 150-nm thick gold interdigitated electrodes with 4 $\mu$m-wide gaps that enable for the application of an electric field. A COMSOL simulation of the geometry of the electrodes indicates electric field magnitudes in the range $[80 , 160]$ kV/cm for an applied voltage of +100 V, depending on the distance from the metal. 

The DBT:Ac nanocrystal growth procedure consists in injecting 100 $\mu$L of a 1:25000 mixture of 1mM DBT-toluene solution and 5 mM Ac-acetone solution into 1 mL of sonicating water. During the 30 min long sonication, solvents dissolve in water and DBT:Ac crystals are formed. The suspension is drop-casted onto the interdigitated electrodes and the water evaporated in vacuum. Solvents and Ac were purchased from Sigma-Aldrich, water was deionized by a Milli-Q Advantage A10 system (18.2 m$\Omega$ cm at 25 °C), and DBT was purchased from Mercachem.

The DBT-doped nanocrystals are dispersed on the electrodes and then covered with a 100 nm layer of polivinyl alcohol (PVA, dissolved in water at 5$\%$ concentration, molecular weight = 13000-23000 purchased from Sigma-Aldrich) via spin-coating (5000 rounds per minute for 120 s), in order to avoid sublimation of the Ac matrix during the cooldown of our cryostat. Fig. \ref{fig:1}(a) shows a sketch of the sample as well as the reference frame used throughout this discussion. The x-axis is the direction of the applied electric field while the z-axis corresponds to the vertical direction and coincides with the optical path of the confocal microscope.

\begin{figure}
    \centering
    \includegraphics[width = 0.5\textwidth]{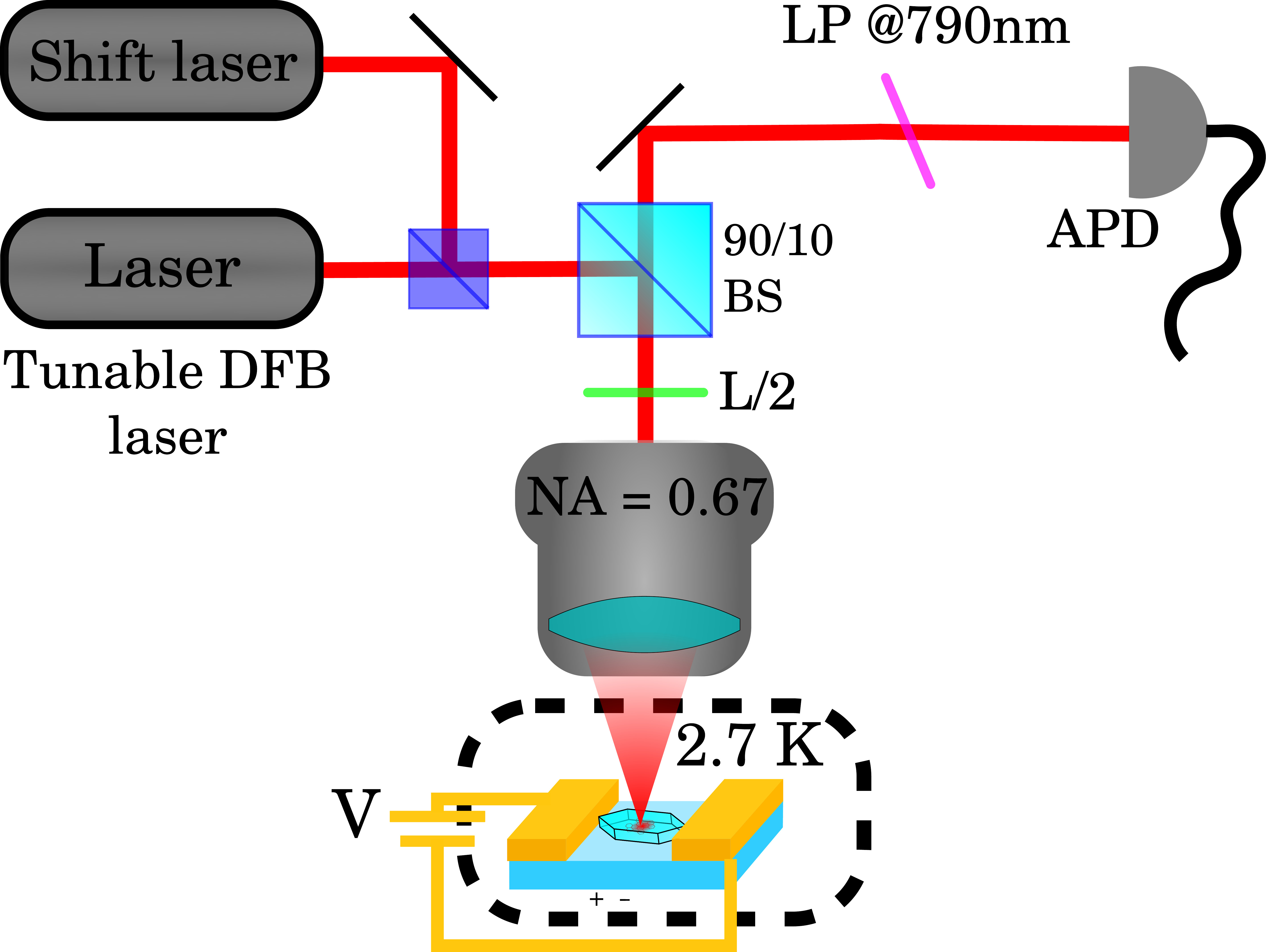}
    \caption{\textbf{Optical Setup.}  Molecules are excited and observed in a confocal microscope setup whose main elements are here schematically shown. BS stands for beam splitter, L/2 for half-wave plate, LP for long-pass filter and APD for avalanche photodiode.} 
    \label{fig:exp setup}
\end{figure}

%The system is studied on a confocal microscope setup where we use excitation spectroscopy to study the emission properties of the DBT molecules \cite{what article do we cite where we explain the setup?}. In particular we use a continuous wave DFB laser (Toptica LD-0785-0080-DFB-1) that can be reproducibly tuned in a range between 783 and 786 nm. By scanning the laser on the ZPL of a molecule, collecting the emission and spectrally isolating the Stokes-shifted fluorescence we measure the shape of the ZPL and its absolute frequency.

A scheme of the optical system used in the experiments is shown in Fig. \ref{fig:exp setup}. The sample is glued on the cold finger of our closed-cycle cryostat (Montana Cryostation) by means of thermally conductive silver paste (RS PRO 
123-9911). In order to achieve the desired thermal contact between the sample and the cold finger, we let the silver paste fully dry before the vacuum pumping of the cold chamber, this procedure avoids the formation of bubbles underneath the sample that would reduce the thermal contact and thus increase the temperature at the molecule position. The DBT molecules, kept at 2.7 K, are excited on their ZPLs using a continuous wave DFB laser (Toptica LD-0785-0080-DFB-1) that can be reproducibly tuned in a range between 783 and 786 nm. A high numerical aperture objective (NA = 0.67, SigmaKoki PAL-50-NIR-HR-LC07) is used to both focus the excitation beam on the emitter of interest and collect the emission from it. A half-wave plate at the back entrance of the objective allows for the selection of the polarization of the excitation light. The laser beam is separated from the detection path using a 90/10 unbalanced beam splitter followed by a longpass filter (Semrock Razoredge-785RS-25), rejecting the scattered laser light. The Stokes-shifted fluorescence emitted by the molecule is transmitted by the filter and its total intensity, directly proportional to the population of the excited state of the ZPL transition, is then focussed on a single photon avalanche photodiode (APD, Excelitas SPCM 800 14 FC). Keeping the power of the excitation laser lower than the saturation intensity of the single emitters and scanning the laser frequency over the ZPL, the collected signal reproduces the shape of the emission line and provides a measurement of the absolute frequency of each DBT molecule. This procedure is what has been dubbed excitation spectroscopy.

The laser-induced shift is implemented by using a continuous wave laser at 767 nm (Toptica DXL 110) focussed on the emitter of interest. Using power in a range between 100 $\mu$W and 1 mW we detect a ZPL shift. The magnitude of the shift is small when compared to previously published results because in this case the observed nanocrystals do not lie on a mirror that would have increased the power density at the confocal spot. Since the optical shift is semi-permanent and persists for hours after switching off the shifting laser, it is possible to follow the shift of the line using excitation spectroscopy between different shift rounds.

\renewcommand{\figurename}{Figure S.}
\setcounter{figure}{0}
\begin{suppinfo}
    \section{The Stark effect in DBT molecules}

We consider here the effect of an external electric field $\vec E_{\rm ext}$ on the spectral properties of a DBT molecule.
In particular we focus on the so-called zero phonon line (ZPL), which here is given by the lowest electronic transition between the ground state $|0\rangle$ and the first singlet excited state $|1\rangle$.
By means of perturbation theory we obtain a shift of the $0\rightarrow 1$ transition frequency in the form of
\begin{equation}
    \hbar\Delta \omega = \vec{d}\cdot \vec E_{\rm ext} - E_{\rm ext}^i\alpha_{ij}E_{\rm ext}^j.
\end{equation}
Here the dipole moment is defined as $\vec{d}=e(\langle{1| \vec{r} |1\rangle} - \langle{0| \vec{r} |0\rangle})$ and it gives the first order of the perturbation theory.
For symmetry reason we can safely set it to zero $\vec{d}\approx 0$.
The second order part instead is given by the polarizability tensor, formally written as
\begin{equation}
    \alpha_{ij} = \frac{e^2}{2}\left[ \sum_{n\neq 1} \frac{\langle{1| r_i |n\rangle}\langle{n| r_j |1\rangle}}{\hbar\omega_n - \hbar\omega_1} - \sum_{n\neq 0} \frac{\langle{0| r_i |n\rangle}\langle{n| r_j |0\rangle}}{\hbar\omega_n - \hbar\omega_0} \right]
\end{equation}
where $|n\rangle$ are the electronic eigenstates of the molecule.
In this treatment we neglect the phononic contributions, which are much out of resonance and they thus bring an almost zero contribution.

Considering the DBT main axis we have that the polarizability tensor is diagonal, completely determined by the $\alpha_{xx}, \alpha_{yy}, \alpha_{zz}$ components.
Anyway, computing the polarizability tensor can be extremely challenging, since it requires to solve the full many-body electronic structure of the molecule.

Fortunately we can still obtain useful information by truncating that expression to the first three level (we notice that two levels are not enough, since the resulting polarizability would be always negative, in contrast with current observations). For each component we need to consider the first three levels for which the dipole matrix element are non-zero. 
For the $xx$ component we can infer this information from a previous work where the electronic structure and main optical properties are explicitly computed\cite{Sadeq2018}.
We then arrive to
\begin{equation}    
\alpha_{xx} \approx \frac{e^2}{2}\left[ \frac{\langle{1| x |2\rangle}\langle{2| x |1\rangle} }{\hbar\omega_2 - \hbar\omega_1} - 2\frac{\langle{0| x |1\rangle} \langle{1| x |0\rangle} }{\hbar\omega_1 - \hbar\omega_0} \right].
\end{equation}
With this expression we can now estimate the polarizability tensor.

The ZPL frequency of the DBT gives $\hbar\omega_1 - \hbar\omega_0\approx 1.6$eV, while we can estimate $\hbar\omega_2 - \hbar\omega_1\approx 2$eV, identifying it as a two-photon transition\cite{Sadeq2018}.
The matrix element $\langle{1| x |0\rangle}$ can be obtained from the DBT ZPL dipole transition, which is typically assumed around $e\langle{1| x |0\rangle} \sim 12$D \cite{Sadeq2018}.
The matrix element of the $1\rightarrow 2$ transition is not known, but we can guess it.
For a harmonic oscillator we would have $\langle{2| x |1\rangle} = \sqrt{2}\langle{1| x |0\rangle}$, for which the polarizability is exactly zero.
So we expect that this matrix element can be assumed larger than this, reasonably still on the same order, giving an estimate $e\langle{2| x |1\rangle} \sim 20-30$D.
Putting everything together we can estimate the following range for the DBT polarizability
\begin{equation}
    \frac{\alpha_{xx}}{\hbar} \approx 2\pi \times 0.2-2 ~ {\rm MHz/(kV/cm)}^2,
\end{equation}
which is in agreement with previous estimations\cite{Nicolet2007a}.

Regarding the $z$-direction polarizability, we can guess to be non zero, but smaller than the $x$-direction one, $\alpha_{zz} < \alpha_{xx}$. This can be deduced again by considering the electronic eigenstates structure described\cite{Sadeq2018}.

\section{Effect of field charge on the excitation spectrum}

The excitation spectrum, observed (or integrated) on time scale comparable or longer than the SD characteristic time, will be described by the convolution of the Fourier-limited DBT Lorentzian line centered at frequency  $\omega$, $\textsl{L}(\omega, \gamma_0)$, with the distribution probability associated to the  fluctuating frequency shift $p( \delta \omega )$  describing SD
\begin{equation}\label{supp-eq:Spectrum_theory_general}
    S(\omega) = \int_{-\infty}^{+\infty} p(\delta \omega)\textsl{L}(\omega-\delta \omega, \gamma_0)d\delta \omega .
\end{equation}
In general, there might be several other mechanisms responsible for spectral diffusion in quantum emitters, resulting in a central frequency which is randomly distributed, with a standard deviation $\sigma$, that is strongly dependent on the observation time. These are typically described in terms of interaction with a bath of two-level systems that can generate different broadening regimes depending on the coupling of the emitter to the bath and to the time scales involved in the process. Different regimes correspond to different $p(\delta\omega)$ following lorentzian, gaussian or Levy statistics for the most common cases \cite{Reilly1994, Gmeiner2016}.
A Gaussian broadening is used to analyze the measurements in this paper, resulting in integrated spectra like in Eq. \eqref{supp-eq:Spectrum_theory_general} with the shape of a Voigt profile. This kind of broadening matches rigorously results obtained from the model when the fluctuation of the electric field are small enough that the paraboloid of Fig. \ref{fig:1}(c) of the main text can be locally approximated with a plane.

\section{Stark effect and spectral diffusion after EGOSS}

\begin{figure}
    \centering
    \includegraphics[width=\textwidth]{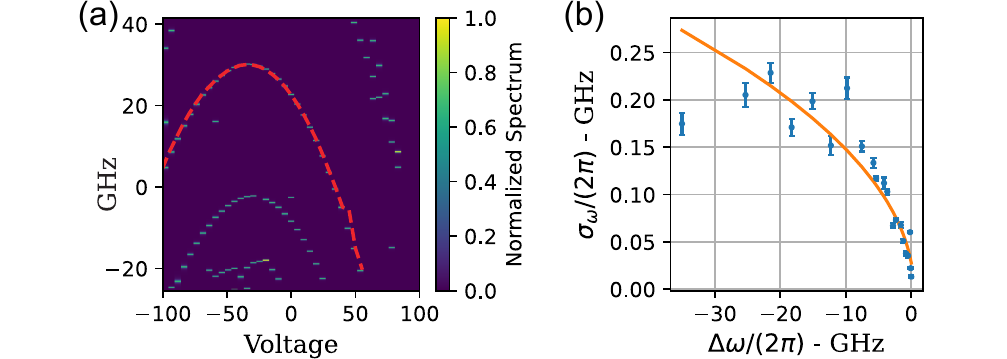}
    \caption{(a) Excitation-spectra $S(\omega)$ map as a function of the applied voltage on the electrodes. This measurement is obtained at a scan speed of 1 GHz/s.
    (b) Measured spectral diffusion as a function of the frequency shift extracted from the data highlighted by the red dashed line in (a). The values of $\sigma_{\omega}$ are extracted by fitting each fixed voltage spectrum with a Voigt profile. }
    \label{supp-fig:7}
\end{figure}

In Fig. S.\ref{supp-fig:7}(a) we show another observation of the Stark effect on a different molecule with respect to Fig. \ref{fig:3}(b).

From the highlighted parabola in Fig.S.\ref{supp-fig:7}(a) we estimate the DBT polarizability $\alpha_{xx}/\hbar  \approx 2\pi \times 1.65 $MHz/(kV/cm)$^2$, still consistent with what presented in the main text.

As in the main text, the excitation spectrum is fitted with a Voigt profile fixing the Lorentzian component on the Fourier limit $\gamma_{\rm lim}/(2\pi)\approx 80$ MHz. The extracted value for $\sigma_{\omega}$ in Fig. S.\ref{supp-fig:7}(b) is purely due to SD. We observe that in this case the scanning speed was the double of that used for the measurement plotted in Fig.\ref{fig:2} and Fig.\ref{fig:3} (total scan time $\bar{t}\sim 1$s). Besides the variability among molecules, this shorter time scales helps explaining the smaller observed SD.
$\sigma_{\omega}$ is fitted as a function of the measured Stark shift $\Delta \omega_{\rm ZPL}$ using Eq. \eqref{eq:sigma_vs_StarkShift}, as is shown in Fig. S.\ref{supp-fig:7}(b), obtaining $\alpha_{xx}\sigma_{E}^2 /\hbar = 2\pi \times 0.53 \pm 0.01$\, MHz. 
Using the estimated polarizability we have that the local electric field variance is $\sigma_E = 1.56\pm 0.20$\,kV/cm.
This value is again compatible with what reported in the main text.

\section{Long term SD time traces}

We here report the time traces obtained for the SD measures whose integrated versions are in the insets of Fig.\ref{fig:5} of the main text.

\begin{figure}
    \centering
    \includegraphics[width=\textwidth]{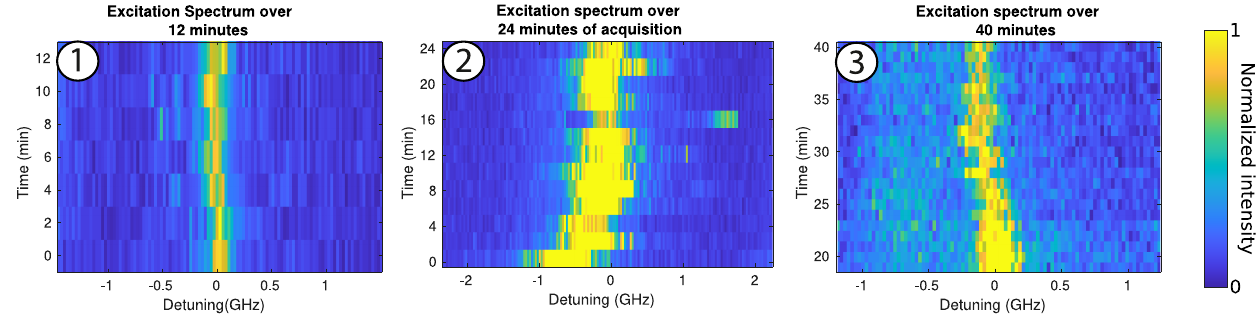}
    \caption{\textbf{SD traces.} The excitation spectrum from a DBT emitter, obtained recording the red-shifted fluorescence signal as a function of the excitation laser frequency scanned at a speed of 0.5 GHz/s, is reported in color scale. The swipes are obtained within 2 s with 2 minutes waiting time between after each one. The three measurements correspond to the conditions described in Fig. \ref{fig:5} where only the integrated versions of the data are shown.}
    \label{supp-fig:8}
\end{figure}

Measured SD induced broadening can strongly depend on the time of observation. Data reported in Fig.\ref{fig:3}(c) and in Fig.S.\ref{supp-fig:7}(c) can be used to validate our model and to extract estimates of the orders of magnitude of relevant quantities like $\sigma_{E}$. However while those measurements were performed over short periods of time (2 s for each point), we monitored SD fluctuations over much longer times having in mind the time scales of experiments similar to those previously performed in our group\cite{Lombardi2021, Duquennoy2022}. In particular measurements 1, 2 and 3 have been integrated over 12, 24 and 42 minutes respectively.
    
\end{suppinfo}

%%%%%%%%%%%%%%%%%%%%%%%%%%%%%%%%%%%%%%%%%%%%%%%%%%%%%%%%%%%%%%%%%%%%%
%% The same is true for Supporting Information, which should use the
%% suppinfo environment.
%%%%%%%%%%%%%%%%%%%%%%%%%%%%%%%%%%%%%%%%%%%%%%%%%%%%%%%%%%%%%%%%%%%%%

\begin{acknowledgement}

This work has been co-funded by the European Union - NextGeneration EU, "Integrated infrastructure initiative in Photonic and Quantum Sciences" - I-PHOQS [IR0000016, ID D2B8D520, CUP B53C22001750006]. The research has been cofunded by the EC under the FET-OPEN- RIA project STORMYTUNE (G.A. 899587)It is also co-funded by the European Union (ERC, QUINTESSEnCE, 101088394). Views and opinions expressed are however those of the author(s) only and do not necessarily reflect those of the European Union or the European Research Council. Neither the European Union nor the granting authority can be held responsible for them. The authors wish to acknowledge Pietro Lombardi for advice on sample fabrication. 

\end{acknowledgement}

%%%%%%%%%%%%%%%%%%%%%%%%%%%%%%%%%%%%%%%%%%%%%%%%%%%%%%%%%%%%%%%%%%%%%
%% The appropriate \bibliography command should be placed here.
%% Notice that the class file automatically sets \bibliographystyle
%% and also names the section correctly.
%%%%%%%%%%%%%%%%%%%%%%%%%%%%%%%%%%%%%%%%%%%%%%%%%%%%%%%%%%%%%%%%%%%%%
%\bibliography{Bibliography_zotero}
\bibliography{publicationsJabRef2}
%\bibliography{Biblioteca personale2}
\end{document}